\documentclass[twocolumn,prl,aps,superscriptaddress,showpacs,amsmath,amssymb,floatfix,shortbibliography]{revtex4-1}
\usepackage{graphicx}
\usepackage{amssymb}
\usepackage{amsmath}
\usepackage{epsfig}
\usepackage{color}
\usepackage{mathtools}
\usepackage[colorlinks,linkcolor=blue,anchorcolor=blue,citecolor=blue,urlcolor=blue]{hyperref}

\begin{document}

\title{Symmetry-enriched Bose-Einstein condensates in spin-orbit coupled bilayer system}
\author{Jia-Ming Cheng}
\affiliation{CAS Key Lab of Quantum Information, University of Science and Technology of China, Hefei, 230026, P.R. China}
\author{Xiang-Fa Zhou}
\affiliation{CAS Key Lab of Quantum Information, University of Science and Technology of China, Hefei, 230026, P.R. China}
\affiliation{Synergetic Innovation Center of Quantum Information and Quantum Physics, University of Science and Technology of China, Hefei, 230026, P.R. China}
\author{Zheng-Wei Zhou}
\email{zwzhou@ustc.edu.cn}
\affiliation{CAS Key Lab of Quantum Information, University of Science and Technology of China, Hefei, 230026, P.R. China}
\affiliation{Synergetic Innovation Center of Quantum Information and Quantum Physics, University of Science and Technology of China, Hefei, 230026, P.R. China}
\email{gongm@ustc.edu.cn}
\author{Guang-Can Guo}
\affiliation{CAS Key Lab of Quantum Information, University of Science and Technology of China, Hefei, 230026, P.R. China}
\affiliation{Synergetic Innovation Center of Quantum Information and Quantum Physics, University of Science and Technology of China, Hefei, 230026, P.R. China}
\author{Ming Gong}
\email{gongm@ustc.edu.cn}
\affiliation{CAS Key Lab of Quantum Information, University of Science and Technology of China, Hefei, 230026, P.R. China}
\affiliation{Synergetic Innovation Center of Quantum Information and Quantum Physics, University of Science and Technology of China, Hefei, 230026, P.R. China}
\date{\today }

\begin{abstract}
We consider the fate of Bose-Einstein condensation (BEC) with time-reversal symmetry and inversion symmetry in a spin-orbit coupled bilayer system. 
When these two symmetry operators commute, all the single particle bands are exactly two-fold degenerate in the momentum space. 
The scattering in the two-fold degenerate rings can relax the spin-momentum locking effect resulting from spin-orbit coupling, thus we can realize the spin 
polarized plane wave phase even when the inter-particle interaction dominates. When these two operators anti-commute, the lowest two bands may have the same 
minimal energy, which have totally different spin structures. As a result, the competition between different condensates in these two energetically 
degenerate rings can give rise to interesting stripe phases with atoms condensed at two or four colinear momenta. We find that the crossover between these two cases is 
accompanied by the excited band condensation when the interference energy can overcome the increased single particle energy in the excited band. This effect is
not based on strong interaction, thus can be realized even with moderate interaction strength.
\end{abstract}


\maketitle

The synthetic gauge potentials in ultracold atoms have attracted much attention in recent years \cite{fu2014production,zhang2013stability,zhang2012collective,leblanc2013direct,bloch2008many,lewenstein2007ultracold,qu2013observation}.  In the Abelian gauge potential the neutral atoms can experience a light-induced commutative vector potential \cite{lin2009bose,lin2009synthetic,beeler2013spin}, thus a lot of interesting phases in charged electrons from Lorentz force can  be simulated using the neutral atoms \cite{jaksch1998cold,greiner2002quantum,greiner2003emergence,bourdel2004experimental,bartenstein2004crossover,zwierlein2004condensation,regal2004observation,zwierlein2003observation,kinast2004evidence}, including the integer and fractional quantum Hall effects \cite{sorensen2005fractional, goldman2009non,baranov2005fractional}.  The non-Abelian potential, such as the spin-orbit coupling (SOC) effect, is more intriguing due to the spin-momentum locking effect. It not only fundamentally changes the fate of Bose-Einstein condensation (BEC) in Bose gases \cite{stanescu2008spin,wang2010spin,cong2011unconventional,hu2012spin,ho2011bose, zhou2013fate,sun2015tunneling,su2016rashba}, but may also give rise to topological phases in Fermi gases \cite{gong2011bcs,hu2011probing,liu2013topological,wu2013unconventional,chen2012bcs}. Thus the synthetic gauge potentials in ultracold atoms have opened a totally new avenue in exploring fundamental physics \cite{jaksch2003creation,celi2014synthetic,gerbier2010gauge,goldman2014light,mittal2014topologically,anderson2012synthetic,hauke2012non,vyasanakere2011bcs,dalibard2011colloquium}; and to this end at least nine experimental groups have realized the SOC in different atoms \cite{lin2011spin,cheuk2012spin,wang2012spin,kolkowitz2016spin,luo2016tunable,li2016spin,ji2014experimental,olson2014tunable,hamner2014dicke,huang2016experimental,wu2016realization}.

The SOC can modify the topology of single particle band from a minimal point at $\mathbf{k}=0$ to a degenerate ring (or sphere) at $\vert\mathbf{k}\vert \ne 0$ in two (or three) 
spatial dimensions (assuming the system has rotational symmetry). In the later case the BEC is impossible to be developed due to the infinite degeneracy of the ground state space.
However, in the presence of interaction, the atoms are forced to occupy only a few momenta. The ground state phase is then determined by the minimized interaction energy, 
which favors either spin balanced plane wave (PW) phase when $c_{12} < 1$ or spin fully polarized PW phase without SOC and standing wave (SW) phase with SOC when $c_{12} > 1$, 
where $c_{12}$ denotes the ratio between inter-particle and intra-particle scattering strengths \cite{wang2010spin}. This general principle has been applied to understand the 
exotic two-component BEC in various circumstances \cite{stanescu2008spin,wang2010spin,cong2011unconventional,hu2012spin,ho2011bose, zhou2013fate,sun2015tunneling,su2016rashba}, in
which the PW phase is almost impossible to be developed when inter-particle interaction dominates.

We find that it is possible to go beyond the above general scenario by considering the BEC in two degenerate rings in a spin-orbit coupled bilayer system.
({\bf I}) When all the bands are two-fold degenerate due to commutation relation between time-reversal operator and inversion operator,
the scattering in the degenerate space can relax the spin-momentum locking effect and spin polarized PW phase can be realized even when $c_{12} > 1$. 
The spin polarization is controlled by the interlayer tunneling. ({\bf II}) When the two operators anti-commute, the condensates in the lowest two energetically degenerate rings with 
distinct spin structures can give rise to different spin balanced colinear stripe (ST) phases occupying two or four momenta; ({\bf III}) The crossover between the above two cases is 
accompanied by the condensation in the excited band due to the interference effect, which can be realized even with moderate interaction strength. 

\begin{figure}[!htb]
\includegraphics[width=2.8in]{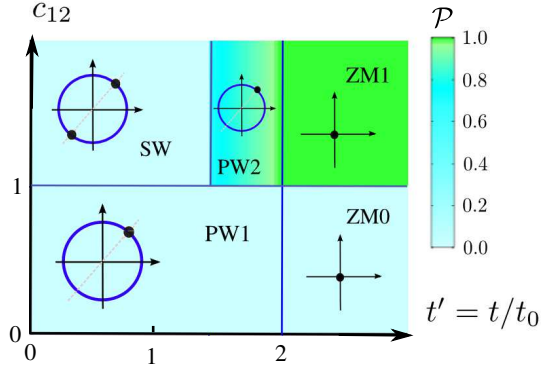}
    \caption{(Color online) Phase diagram for BEC in two identical degenerate rings. Here SW, PW and ZM are short for standing wave, plane wave and zero momentum wave, respectively.
    In each phase the condensate is denoted by solid cycle(s) and $\mathcal{P} = |\langle \sigma_z\rangle|$ denotes the corresponding spin polarization.}
    \label{fig-phasediag1}
\end{figure}

\textit{Model and Hamiltonian}. We consider a bilayer system with Rashba SOC \cite{wu2016realization}. Along the perpendicular direction, the tunneling between 
the two layers are assumed to be spin-dependent $t_{\uparrow} = -t_{\downarrow} = t$, which can be realized by fast modulating the optical lattice \cite{lignier2007dynamical, zenesini2009coherent, 
meinert2016floquet}. The single particle Hamiltonian reads as $H_{0}=\int d\mathbf{r}\psi^{\dagger}(\mathbf{r})\mathcal{H}_{0}\psi(\mathbf{r})$, with
$\mathcal{H}_{0}= \frac{\mathbf{k}^{2}}{2m} + \frac{\lambda}{m}\sigma_0\otimes\mathbf{k}\cdot\boldsymbol{\sigma}-\sigma_x\otimes\tilde{t}$ and $\tilde{t} =\text{diag}\{t_{\uparrow},t_{\downarrow}\}$,
under the basis $\psi(\mathbf{r})=(\psi_{1\uparrow}(\mathbf{r}),\psi_{1\downarrow}(\mathbf{r}),\psi_{2\uparrow}(\mathbf{r}),\psi_{2\downarrow}(\mathbf{r}))^{T}$.  Here $\lambda$ is the SOC strength, 
${\bf k} = (k_x, k_y)$, and $\boldsymbol{\sigma} = (\sigma_x, \sigma_y)$ are Pauli matrices. Notice that the spin-dependent tunneling introduces a relative $\pi$ phase 
between the two spin components, which is equivalent to introduce a minus sign to the SOC coefficients between the two layers. Thus the above model can be mapped to the following model by a 
unitary rotation,
\begin{eqnarray}
  \mathcal{H}_{0}= \frac{\mathbf{k}^{2}}{2m} + \frac{\lambda}{m}\sigma_z\otimes\mathbf{k}\cdot\mathbf{\sigma}-t\sigma_x\otimes\sigma_0,
  \label{eq-H0}
\end{eqnarray}
for $\psi(\mathbf{r})=(\psi_{1\uparrow}(\mathbf{r}),\psi_{1\downarrow}(\mathbf{r}),\psi_{2\uparrow}(\mathbf{r}), -\psi_{2\downarrow}(\mathbf{r}))^{T}$. We neglect the
interlayer interaction, thus the interacting Hamiltonian can be written as \cite{sun2015tunneling,su2016rashba},
\begin{equation}
    \mathcal{V}_{\text{int}}=\frac{g}{2}\sum_{i=1,2}\int d\mathbf{r} [n_{i}^{2}(\mathbf{r})+2(c_{12}-1)n_{i\uparrow}(\mathbf{r})n_{i\downarrow}(\mathbf{r})],
    \label{eq-Hint}
\end{equation}
where $n_{i\sigma}(\mathbf{r})=\psi_{i\sigma}^{\dagger}(\mathbf{r})\psi_{i\sigma}(\mathbf{r})$, $n_i(\mathbf{r})=\sum_{\sigma}n_{i\sigma}(\mathbf{r})$, and $c_{12}$ 
defines the ratio between inter-particle and intra-particle interaction strengths. In the following, Eq. \ref{eq-H0} is used in our simulation; and the total Hamiltonian $\mathcal{H} = \mathcal{H}_0 + \mathcal{V}_\text{int}$.

This model is intriguing due to the presence of time-reversal symmetry $\mathcal{T}$ and  inversion symmetry $\mathcal{I}$ between the two layers defined as \cite{morimoto2013topological}
\begin{equation}
  \mathcal{T}=\sigma_0\otimes i\sigma_y K, \quad \mathcal{I}=\sigma_x\otimes\sigma_0,
    \label{eq-IS}
\end{equation}
where $K$ is the complex conjugate operator. The combination of the above two operators enables the definition of a new anti-unitary operator, $T = \mathcal{I}\cdot \mathcal{T} = \sigma_x \otimes i \sigma_y K$, 
which satisfies,
\begin{equation}
    T \mathcal{H}_0({\bf k}) T^{-1} = \mathcal{H}_0({\bf k}), \quad T^2 = -1.
    \label{eq-T}
\end{equation}
The above result is based on the fact that $\mathcal{I}$ and $\mathcal{T}$ commute, {\it i.e.}, $[\mathcal{I}, \mathcal{T}] = 0$. 
The Kramers' theorem ensures that all the bands at each ${\bf k}$ are 
exactly two-fold degenerate. In the single layer system \cite{wang2010spin}, these two symmetry operators anti-commute, thus the model does not have 
the above feature. For the model in Eq. \ref{eq-H0}, we have
\begin{eqnarray}
	\epsilon^{\pm}(\mathbf{k})=t_0[(\frac{ k}{\lambda})^{2}\pm(4(\frac{ k}{\lambda})^{2}+t^{\prime2})^{1/2}],
\end{eqnarray}
with $t^{\prime}=t/t_0$ and $t_0=\lambda^2/(2m)$. Hereafter $t_0$ is used as the basic energy scale throughout this work. The topology of the ground state space is controlled 
by $t'$, and the degenerate rings with radius $k = \lambda \sqrt{1-t'^2/4}$ will shrink to a point at ${\bf k} = 0$ when $t' \ge 2$. When $t' > 2$, the two 
degenerate eigenvectors at ${\bf k} = 0$ read as
\begin{equation}
    \phi_{1}^{\eta}({\bf r})= (1,\eta ,1, \eta)^T/2,  \quad \eta = \pm,
    \label{eq-gs1}
\end{equation}
with single particle energy $\epsilon^{-}_{\text{min}}= -t't_0$. When $t' < 2$, the wave functions at the degenerate rings are,
\begin{equation}
    \phi_{2\mathbf{k}}^\eta({\bf r}) =
	(\gamma_\eta e^{-i\theta_{\mathbf{k}}}, \eta \gamma_\eta, \gamma_{-\eta} e^{-i\theta_{\mathbf{k}}}, \eta \gamma_{-\eta})^T e^{i{\bf k}\cdot {\bf r}}/2,
    \label{eq-gs2}
\end{equation}
with $\gamma_{\eta}=(1-\eta(1-t^{\prime2}/4)^{1/2})^{1/2}$, $e^{i\theta_{{\bf k}}} = (k_x + ik_y)/k$ and $\epsilon^{-}_{\text{min}}=-t_0 (1+t'^2/4)$.
These wave functions are normalized via $\int d\mathbf{r}\vert\phi^{\eta}_{1}(\mathbf{r})\vert^2=\int d\mathbf{r}\vert\phi^{\eta}_{2\mathbf{k}}(\mathbf{r})\vert^2=1$. 

\begin{figure}[!htb]
\includegraphics[width=2.8in]{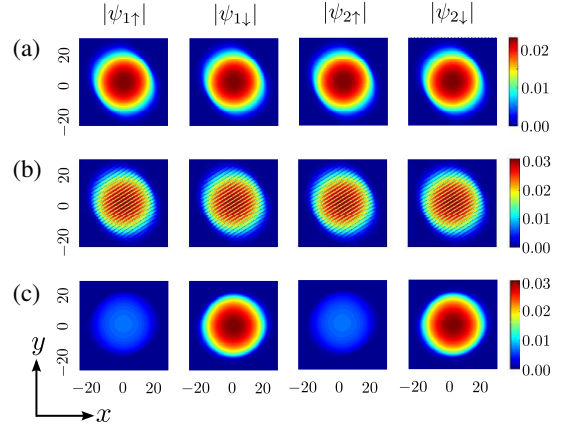}
    \caption{(Color online) Ground state densities in a trap for (a) PW1 phase with ($t', c_{12}$)=(1.0,0.8), (b) SW phase with ($1.0, 1.2$) and (c) PW2 phase with (1.8, 1.2). 
    In imaginary time-evolution simulation, $\lambda=2$, $g=100$, $\delta=-1$, $m=\hbar=1$.}
    \label{fig-wf}
\end{figure}

\textit{Phase Diagram}. We combine the variational analysis and imaginary time-evolution method (confined in a weak trap) to determine the fate of BEC
by minimizing the total energy per particle, $\mathcal{E}_g = \mathcal{E}_0 + \mathcal{E}_\text{int}$, where $\mathcal{E}_0 =\int d\mathbf{r} \phi^{\ast}(\mathbf{r})\mathcal{H}_{0}
\phi(\mathbf{r})=\epsilon^{-}_{\text{min}}$ denotes the single particle energy \cite{ueda2010fundamentals}. The interaction is essential for the development of BEC over a degenerate space via the order-by-disorder 
mechanism, which singles out a ground state from the degenerate space. We find that when more than one momenta are occupied, all these momenta should be collinear. 
These two methods yield the same phase diagram as shown in Fig. \ref{fig-phasediag1} as a function of tunneling and relative interaction strength.

Firstly we consider the case when all the atoms are condensed at $\mathbf{k}=0$, in which the wave function should be a superposition of the two 
orthogonal states in Eq. \ref{eq-gs1}. We find (a) when $c_{12}<1$, the ground state is a spin balanced phase (denoted as ZM0) in Fig. \ref{fig-phasediag1},
\begin{eqnarray}
	\psi_{\text{ZM0}}(\mathbf{r})=|\alpha|\phi^{+}_{1}(\mathbf{r})\pm i |\beta|\phi^{-}_{1}(\mathbf{r}),
\end{eqnarray}
with $\mathcal{E}_{\text{ZM0}}=-t + g(1+c_{12})/8$ for arbitrary $\alpha$ and $\beta$. Therefore all components have the same weight, thus
$\mathcal{P} = \vert \langle \psi_{\text{ZM0}} | \sigma_z | \psi_{\text{ZM0}}\rangle \vert = 0$. By contrast in case (b) when $c_{12}>1$, 
the ground state is fully spin polarized (denoted as ZM1 in Fig. \ref{fig-phasediag1}),
\begin{eqnarray}
	\psi_{\text{ZM1}}(\mathbf{r})=[\phi^{+}_{1}(\mathbf{r})\pm \phi^{-}_{1}(\mathbf{r})]/\sqrt{2},
\end{eqnarray}
in which $\mathcal{P} =  \vert\langle \sigma_z\rangle\vert =1$ and $\mathcal{E}_{\text{ZM1}}=-t+g/4$. The transition between ZM0 and ZM1, $\mathcal{E}_{\text{ZM1}} = \mathcal{E}_{\text{ZM0}}$, 
determines the boundary at $c_{12} = 1$, across which it is a first-order transition due to discontinuous of $\mathcal{E}$ with respect to $c_{12}$.

Next we turn to the two-fold degenerate rings with $t^\prime < 2$. When $t^\prime \rightarrow 0$, we can recover the results in previous literatures where a first-order phase transition from a PW
phase (denoted as PW1) to SW phase (denoted as SW) is expected at $c_{12} = 1$ (see Fig. \ref{fig-wf}a). Without extra degeneracy, the spin-momentum locking effect
prohibits tuning of spin polarization for a given ${\bf k}$, thus when $c_{12} > 1$, SW phase should be favored to reduce the inter-particle energy. 
This locking effect can be relaxed in the degenerate rings due to the possible scattering between the two states in Eq. \ref{eq-gs2} in the degenerate space, 
where the spin polarization in the PW phase can still be tuned in a wide range by superposition principle.
For instance when $t^\prime \rightarrow 2^{-}$, a spin polarized state ($\mathcal{P} = \vert\langle \sigma_z\rangle \vert \rightarrow 1$)
can be realized by a basis rotation. Thus the PW phase (denoted as PW2) is still allowed when $c_{12} > 1$.

The PW2 phase beyond the general scenario can be understood by minimizing the total energy using the two wave functions in Eq. \ref{eq-gs2}. We find that when 
$c_{12}<1$, the wave function takes the form,
\begin{eqnarray}
    \psi_{\text{{\footnotesize PW1}}}(\mathbf{r})=[\phi^{+}_{2\mathbf{k}}(\mathbf{r})\pm i\phi^{-}_{2\mathbf{k}}(\mathbf{r})]/\sqrt{2},
\end{eqnarray}
with energy $\mathcal{E}_{\text{PW1}}=g(1+c_{12})/8-t_0(1+t^{\prime 2})/4$, and $\mathcal{P} = 0$, thus this phase is spin balanced. However when $c_{12}>1$, 
the wave function reads as
\begin{eqnarray}
	\psi_{\text{PW2}}(\mathbf{r})=[\phi^{+}_{2\mathbf{k}}(\mathbf{r})\pm \phi^{-}_{2\mathbf{k}}(\mathbf{r})]/\sqrt{2},
        \label{eq-PW2}
\end{eqnarray}
with $\mathcal{E}_{\text{PW2}}=g[(1+c_{12})+(1-c_{12})t^{\prime 2}/4]/8-t_0(1+t^{\prime 2}/4)$, and spin polarization $\mathcal{P} = t'/2$ (see the wave functions 
in Fig. \ref{fig-wf}c from imaginary time-evolution simulation). 

In the SW phase, the wave function should be superposition of all the possible waves (Eq. \ref{eq-gs2}) with momenta ${\bf k}_1$ and 
${\bf k}_2$. By direct numerical minimization we find,
\begin{eqnarray}
	\psi_{\text{SW}}(\mathbf{r})=[\psi_{\text{PW1}}(\mathbf{r}) + e^{i\vartheta}\mathcal{T}\psi_{\text{PW1}}(\mathbf{r})]/\sqrt{2},
\end{eqnarray}
with $\mathcal{E}_{\text{SW}}=g^\prime[(1+c_{12})+(1-c_{12})/2]/8-t_0(1+t^{\prime 2}/4)$; see Fig. \ref{fig-wf}b. The phase boundary between SW and PW2 phases is determined by 
$\mathcal{E}_{\text{SW}} = \mathcal{E}_{\text{PW2}}$, which yields $t' = \sqrt{2}$. Thus we have the full phase diagram in Fig. \ref{fig-phasediag1}.

\begin{figure}[!htb]
  \quad
  \includegraphics[width=2.8in]{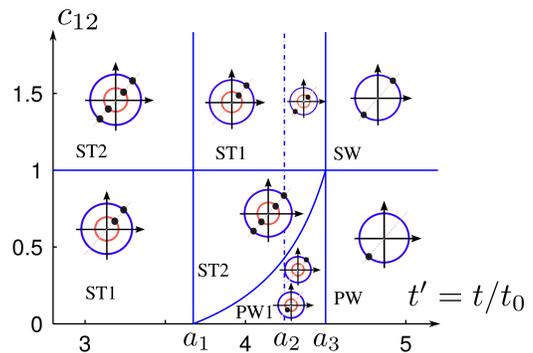}
  \caption{(Color online) Phase diagram for BEC in two energetically degenerate rings when $\lambda_2=\delta \cdot \lambda_1$ ($\delta<0$). Here SW, PW, ST are short for 
  standing wave, plane wave and stripe wave, respectively. $a_{1}=(1-\delta)^{2}/\sqrt{6}, a_{2}=\sqrt{-\delta(1-\delta)^{2}}, a_{3}=(1-\delta)^{2}/2$. In all phases, the 
    spin polarization $\mathcal{P} = 0$.}
    \label{fig-phasediag2}
\end{figure}

\textit{BEC in Two Energetically Degenerate Rings}. We now generalize this idea to another intriguing condition when the two layers have slightly different SOC strengths, say $\lambda_1 = \lambda$
and $\lambda_2/\lambda_1 = \delta$. We still work in the rotated picture in Eq. \ref{eq-H0}, thus $\delta = -1$ corresponds to identical rings discussed in Fig. \ref{fig-phasediag1}. 
In this case the two identical rings will be divided into two rings. We consider
\begin{equation}
    \mathcal{H}_0' = \begin{pmatrix}
        {({\bf k} + \lambda_1 \boldsymbol{\sigma})^2 \over 2m} - {\lambda_1^2 \over 2m} & -t \\
        -t  & {({\bf k} + \lambda_2 \boldsymbol{\sigma})^2 \over 2m} - {\lambda_2^2 \over 2m}
    \end{pmatrix}.
\end{equation}
This model admits a different inversion symmetry, $\mathcal{I}' = \sigma_0 \otimes\sigma_z$. With the method in Eq. \ref{eq-T}, we can define another
anti-unitary operator $T' = \mathcal{I}' \cdot \mathcal{T} = \sigma_0 \otimes \sigma_x K$, with 
\begin{equation}
    T' \mathcal{H}_0'({\bf k}) T'^{-1} = \mathcal{H}_0'({\bf k}), \quad T'^2 = +1.
\end{equation}
Here, $\mathcal{I}'$ anti-commutes with $\mathcal{T}$, {\it i.e.}, $\{\mathcal{I}', \mathcal{T}\} = 0$, similar to that in the single layer system. 
This symmetry means that for any eigenvector, $\mathcal{H}_0' \phi = E \phi$, $T' \phi$ should also be its eigenvector with the same eigenenergy. 
The uniqueness of the solution requires $T' \phi = \phi$ upon a global phase difference, which ensures that all the wave functions should be fully 
spin balanced, {\it i.e.}, $\mathcal{P} = |\langle \phi | \sigma_z| \phi\rangle| = 0$. By a proper unitary rotation we find
\begin{equation}
    \mathcal{H}_0' \rightarrow 
    \begin{pmatrix}
        \epsilon(k_+ + \delta \lambda \sigma_z) - t \sigma_x &  0 \\
        0                                                & \epsilon(k_- + \delta \lambda \sigma_z)- t \sigma_x
    \end{pmatrix},
\end{equation}
where $\epsilon(k) = k^2/(2m)$, $k_{\pm} = k + (\lambda_1 + \lambda_2)/2$ and $\delta\lambda = (\lambda_1 - \lambda_2)/2$.
We find that the upper block and the lower block are identical upon a momentum shift $k \rightarrow k- (\lambda_1 + \lambda_2)$, thus they 
should be degenerate in their eigenvalues upon the same momentum shift; The lowest two bands are 
\begin{eqnarray}
	\epsilon^i(\mathbf{k})=t_0[q^2_i+(1-\delta)^2/4-\sqrt{(1-\delta)^2q^2_i+t^{\prime 2}}],
\end{eqnarray}
with $i = 1, 2$ and $q_i = k/\lambda + (-1)^i (1+\delta)/2$. The time-reversal symmetry ensures that 
$\epsilon^i(k) = \epsilon^i(-k)$ and the momentum translation ensures $\epsilon^1(k) = \epsilon^2(k -(\lambda_1 + \lambda_2))$, thus the two 
rings should have the same minimal energy. This new degeneracy leads to strong competition between different possible 
condensates. The corresponding phase diagram in this degenerate space is presented in Fig. \ref{fig-phasediag2}.

The topology of the single particle bands is controlled by the parameter $\chi = (1-\delta)^{2}-2t^{\prime}$, which have two energetically degenerate rings when $\chi > 0$; otherwise, 
it is a single ring. (a) When $\chi > 0$, the two rings have radiuses $k_{\pm}=[\pm (1+\delta)/2+(1-\delta)(1-4t^{\prime2}/(1-\delta)^{4})^{1/2}/2]\lambda$, with the same 
single particle energy $\epsilon_{\text{min}}(\mathbf{k})=\epsilon^{1}_{\text{min}}(\mathbf{k}_{1})= \epsilon^{2}_{\text{min}}(\mathbf{k}_{2})=t_0[-t^{\prime 2}/(1-\delta)^2]$. The corresponding wave functions for these 
two rings are
\begin{align}
    \label{eq:phik}
    \varphi_{{\bf k}_1}& =(x_{+}(e^{i\theta_{\mathbf{k}_{1}}},-1),x_{-}(e^{i\theta_{\mathbf{k}_{1}}},-1))^Te^{i\mathbf{k}_{1}\cdot\mathbf{r}}, \\ 
    \varphi_{{\bf k}_2}& =(x_{-}(e^{i\theta_{\mathbf{k}_{2}}},1),x_{+}(e^{i\theta_{\mathbf{k}_{2}}},1))^Te^{i\mathbf{k}_{2}\cdot\mathbf{r}}.
\end{align}
where $x_{\mp}=(1\mp(1-4t^{\prime2}/(1-\delta)^{4})^{1/2})^{1/2}/2$. Obviously, $T'\varphi_{{\bf k}_i} = \varphi_{{\bf k}_i}$ and $\mathcal{P} = 0$.
The fate of BEC depends strongly on the scattering between and in the rings, and the final condensates may occupy only
one momentum, two or four momenta in one ring or two rings. We calculate the energies of all these possible phases and minimize these energies 
to determine their boundaries. We find three interesting cases. (1) For the PW phase (PW1),
\begin{eqnarray}
    \psi_{\text{PW1}} = \varphi_{{\bf k}_1} \quad \text{or} \quad\psi_{\text{PW1}} = \varphi_{{\bf k}_2},
\end{eqnarray}
the corresponding total energy is $\mathcal{E}_{\text{PW1}} =-t_0t^{\prime 2}/(1-\delta)^2+ g(1+c_{12})(1/4-t^{\prime 2}/(2(1-\delta)^4))$. 
(2) For the stripe (ST1) phase with two different collinear momenta (${\bf k}_1//{\bf k}_2$), the wave function can be written as,
\begin{eqnarray}
    \psi_{\text{ST1}}=(\varphi_{{\bf k}_1}+\exp(i\vartheta)\varphi_{{\bf k}_2})/\sqrt{2},
\end{eqnarray}
where $\mathcal{E}_{\text{ST1}} = -t_0t^{\prime 2}/(1-\delta)^2+g[(1+c_{12})/8+(1-c_{12})t^{\prime 2}/(4(1-\delta)^4)]$. (3) 
When both rings are occupied by four momenta $\{\mathbf{k}_{1},\mathbf{k}_{2},-\mathbf{k}_{1},-\mathbf{k}_{2}\}$ for the ST2 phase, the wave function can be written as
\begin{equation}
	\psi_{\text{ST2}} = (\psi_{\text{ST1}}+e^{i\vartheta^{\prime}}\mathcal{T}\psi_{\text{ST1}})/\sqrt{2},
\end{equation}
where $\mathcal{E}_{\text{ST2}}=\mathcal{E}_\text{{ST1}}+g(1-c_{12})(1/4-
3t^{\prime 2} /(2(1-\delta)^4))/4$. The SW phase in Fig. \ref{fig-phasediag1} is absent here when $\chi > 0$ due to its slightly higher energy than these three cases. 
These minimized energies are used to determine
the phase diagram presented in Fig. \ref{fig-phasediag2} for $\chi < 0$ (for $t' < a_3$). The boundary between PW1 and ST2 is determined by $2(3+5c_{12})t'^2  = (1+3c_{12})(1-\delta)^4$, 
and the boundary between ST1 and ST2 is determined by $t' = a_1$ and $c_{12} = 1$. Notice that for $t' \in (a_2, a_3)$, the ground state may have two different 
condensates with the same energy. 

\begin{figure}
    \centering
    \includegraphics[width=0.48\textwidth]{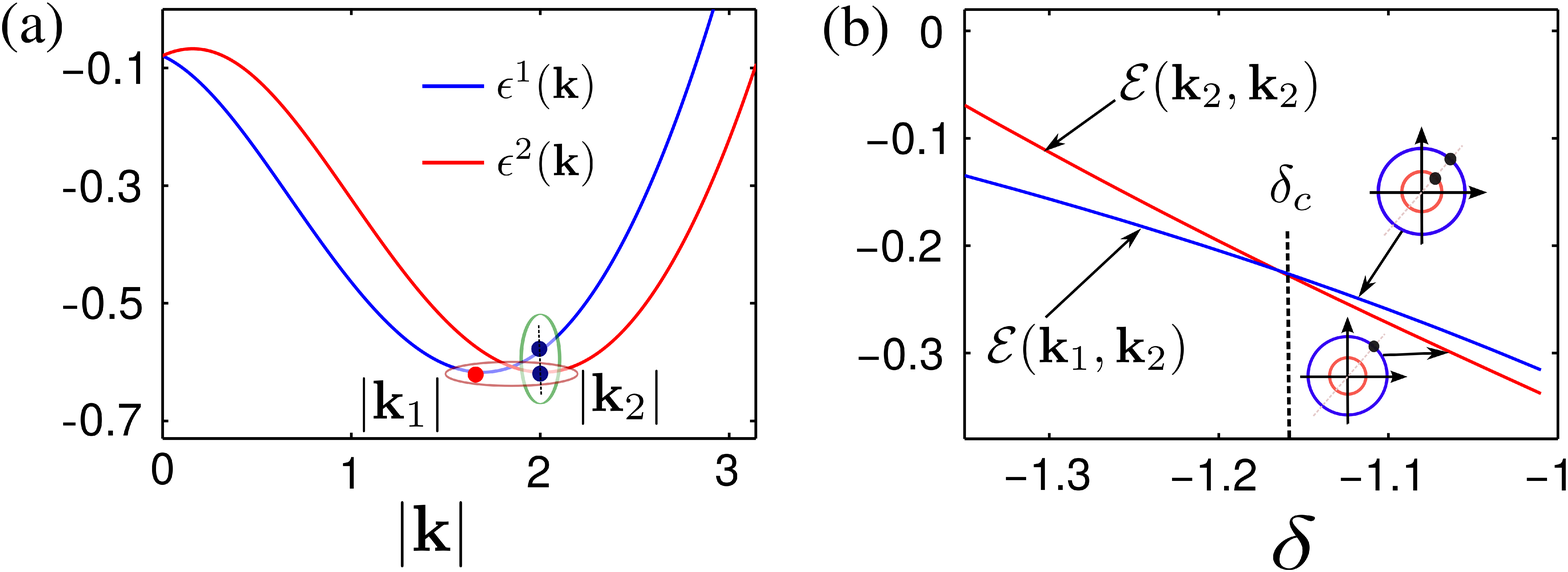}
    \caption{(Color online) (a) The lowest two single particle bands for $\delta = -1.16$. The atoms can occupy either two rings or the ground state and first excited band in a single ring. 
    (b) Ground state energies for these two different condensates. Parameters are: $\lambda = 2$, $t' = 1.2$, $g =2$, $c_{12} = 0.5$ and $\delta_c = -1.16$.}
    \label{fig-delta}
\end{figure}

In case (b) when $\chi \le 0$ the ground state is only made by a single ring. The minimal energy is determined by 
$\epsilon^{2}_{\text{min}}(\mathbf{k})=t_0[(1-\delta)^{2}/4-t^{\prime}]$ at $k=-(1+\delta)\lambda/2$, and its wave function is
\begin{eqnarray}
	\varphi_{{\bf k}}=(e^{i\theta_{\mathbf{k}}},1,e^{i\theta_{\mathbf{k}}},1)^Te^{i\mathbf{k}\cdot\mathbf{r}}/2.
\end{eqnarray}
For the PW phase, the total energy is $\mathcal{E}_\text{PW} = t_0[(1-\delta)^{2}/4-t^{\prime}]+g(1+c_{12})/2$, and for the SW phase, 
\begin{equation}
    \psi_{\text{SW}}= (\varphi_{{\bf k}}+e^{i\theta}\mathcal{T}\varphi_{{\bf k}})/\sqrt{2},
\end{equation}
with $\mathcal{E}_\text{SW} = \mathcal{E}_{\text{PW}}+g(1-c_{12})/4$. We find PW phase is favored when $c_{12} < 1$ and otherwise SW phase, similar to that reported in 
previous literatures \cite{stanescu2008spin,wang2010spin,cong2011unconventional,hu2012spin,ho2011bose,
zhou2013fate,sun2015tunneling,su2016rashba}.

These phases in two separate rings can be understood from the result in degenerate rings in Fig. \ref{fig-phasediag1} by adiabatically switching $\delta$ off from $\delta = -1$, in which 
each condensate in momentum ${\bf k}$ may split into two condensates with different momenta ${\bf k}_{\pm}$. This explains excellently most the phases in Fig. \ref{fig-phasediag1} 
to Fig. \ref{fig-phasediag2} except the ST2 phase and the PW1 phase. In the PW1 phase, only one momentum ${\bf k}$ is allowed due to the reason discussed in Fig. \ref{fig-phasediag1}; 
while in the ST2 phase the competition between different possible condensates is more complicated. This regime has a special physics consequence --- excited band condensation, which 
will be discussed below.

\textit{Excited Band Condensation from Interference Effect}. Notice that $\lim_{\delta \rightarrow -1} a_{2,3}  \rightarrow 2$, thus these two boundaries can be adiabatically tuned to 
the phase boundary at $t'=2$ in Fig. \ref{fig-phasediag1}. However we find, $\lim_{\delta \rightarrow -1} a_1\rightarrow 4/\sqrt{6} \ne \sqrt{2}$, thus the boundary controlled by $a_1$ 
can not be tuned to the phase boundary of SW and PW2 phases in Fig. \ref{fig-phasediag1}. This is attributed to the condensate in the excited band from the interference effect, which can 
be realized even for moderate interaction strength. To this end, let us assume the wave function to be
\begin{equation}
            \psi=\alpha\varphi^{\prime}_{1\mathbf{q}_1}+\beta\varphi^{\prime}_{2\mathbf{q}_2},
\end{equation}
where $\mathbf{q}_1$ and $\mathbf{q}_2$ are not necessary to be restricted to the degenerate rings any more. We find that 
\begin{equation}
    \mathcal{E}(\mathbf{q}_1,\mathbf{q}_2)= T + V + (1 - c_{12}) \mathcal{A} ((\alpha^* \beta)^2  + h.c) \delta_{{\bf q}_1, {\bf q}_2}.
\end{equation}
where $T = |\alpha|^2 \epsilon^1 ({\bf q}_1) + |\beta|^2 \epsilon^2 ({\bf q}_2)$ represents the kinetic energy, $\mathcal{A}$ is a constant depending 
on the details of the wave function, and $V$ is the interacting energy discussed before. The last term, arises from the 
interference effect, is most relevant to our discussion here. When ${\bf q}_1 = {\bf q}_2$, the interference between the particles can further lower 
the total energy by letting $\alpha^*\beta \simeq i/2$ when $c_{12} < 1$, and $\alpha^*\beta \simeq 1/2$ when $c_{12} > 1$.  This lowered energy may 
overcome the increased single particle energy, $|\beta|^2 (\epsilon^2 ({\bf q}_2) - \epsilon^1 ({\bf q}_1))$, when condensed partially in the excited band (see 
Fig. \ref{fig-delta}a). We find that the outer ring always has energy lower than the inner ring, thus let 
\begin{equation}
            \mathcal{E}(\mathbf{k}_1,\mathbf{k}_2)=\mathcal{E}(\mathbf{k}_2,\mathbf{k}_2),
\end{equation}
we are able to determine the critical boundary between condensates in degenerate rings and excited band as
\begin{equation}
            \delta_c=-1+(-B-(B^2-2AB)^{1/2})/(2A),
\end{equation}
with $A=[5(c_{12}-1)gt^{\prime 2}-8(t^{\prime 2}-4)\lambda^2]/128$ and $B=(c_{12}-1)gt^{\prime 2}/32$. For the data in Fig. \ref{fig-delta}b, the predicted boundary for crossover
between the ground state condensate and excited band condensation is $\delta_c=-1.1615$, which is in excellent agreement with the numerical result in Fig. \ref{fig-delta}b. When 
$\delta = -1$, the excited band condensate automatically evolves to the phase diagram in Fig. \ref{fig-phasediag1}. Note that in Fig. \ref{fig-delta}b, $\mathcal{E}(\mathbf{k}_1,\mathbf{k}_2)$,
which does not include the interference term, will not approach $\mathcal{E}(\mathbf{k}_2,\mathbf{k}_2)$ even when $\delta = -1$. This condensate effect is totally different 
from the condensate in excited states, which requires very strong repulsive interaction\cite{zhou2011interaction,zheng2014strong}. 
The regime for excited band condensation can be greatly enhanced by increasing the scattering  strength, thus in principle 
can be observed in a wide range. We also notice that $\mathcal{P} = 0$ when $c_{12} < 1$ and $\mathcal{P} \ne 0$ when 
$c_{12} > 1$ from the results in Fig. \ref{fig-phasediag1}, which provides important basis for experimental detection of the excited band condensation.

\textit{Discussion and Conclusion}. Symmetries play an essential role in condensed matter physics, including the different topological phases classified by time-reversal, particle-hole and 
chiral symmetries \cite{schnyder2008classification,ryu2010topological}. We demonstrate in this work that the time-reversal symmetry and inversion 
symmetry can introduce some extra degeneracy to the single particle band structure in an artificial bilayer system, which can fundamentally change the fate of BEC in the degenerate 
space. These results provide a new route for the realization of exotic BEC using the state-of-art techniques in ultracold atoms.
% It is necessary to point out that these condensates are not protected by symmetries in the sense of topology, thus they will not be immediately destroyed upon weak symmetry breaking.

\begin{acknowledgments}
{\it Acknowledgement}. M.G. is supported by the National Youth Thousand Talents Program (No. KJ2030000001), the USTC start-up funding (No. KY2030000053) and the CUHK RGC Grant (No. 401113). X. Z., Z. Z. and G. G. are supported by National Key Research and Development Program (No. 2016YFA0301700), National Natural Science Fundation of China(No. 11574294), and the ''Strategic Priority Research Program (B)'' of the Chinese Academy of Sciences (No. XDB01030200).
\end{acknowledgments}

\bibliography{Bilayer_soc}

\end{document}